\documentclass{moriond}

\def\be{\begin{equation}}
\def\ee{\end{equation}}
\def\bea{\begin{eqnarray}}
\def\eea{\end{eqnarray}}

\newcommand{\Photo}{\includegraphics[height=35mm]{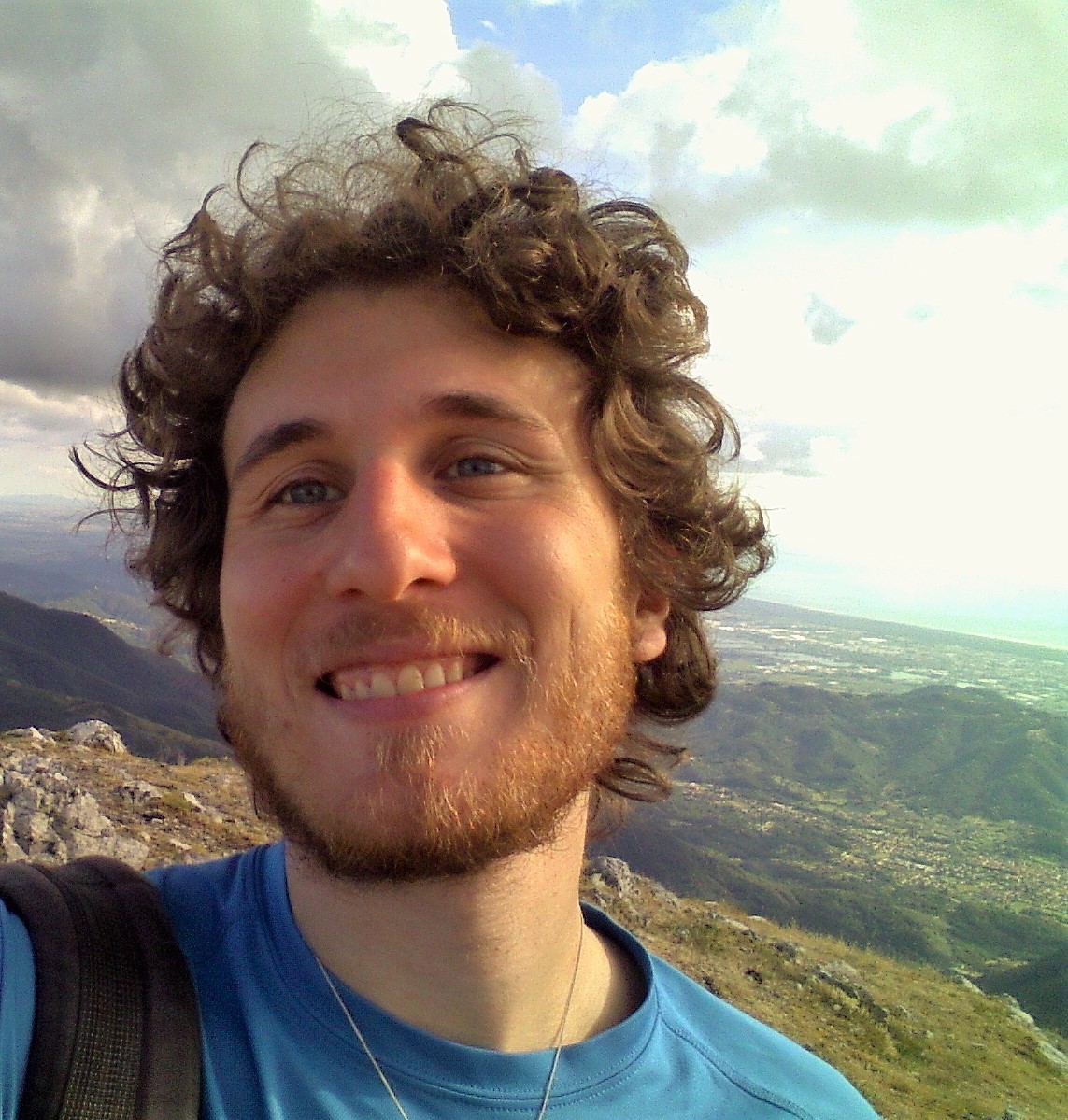}}

\usepackage{hyperref}

\begin{document}
\vspace*{4cm}

\title{Time-dependent, hadronic B decays and electroweak penguins at Belle II}

\author{ Valerio Bertacchi, on behalf of Belle~II Collaboration }

\address{Aix Marseille Univ, CNRS/IN2P3, CPPM, Marseille, France}

\maketitle\abstracts{
We report the recent measurements performed using the data sample collected from 2019 to 2022 by the Belle~II experiment~\cite{Belle2:TDR} at $\Upsilon(4S)$ resonance, corresponding to an integrated luminosity of $362~\mathrm{fb}^{-1}$ . We present the measurement of $B$ lifetime and mixing frequency,  the time dependent CP-violation analyses using the $B^0\to J\psi K_S^0$,  $B^0\to \phi K_S^0$, $B^0\to K_S^0K_S^0K_S^0$, $B^0\to K_S^0\pi^0$ decays  and the implication of the latter for the isospin sum rule,  the observation of $B\to D^{(*)} K^-K_S^0$ decays, the  analysis of $B\to X_s\gamma$. We also present the measurements of the angle $\gamma$ using $B\to DK$ decays, with the combined dataset collected by Belle and Belle II.
}

\section{$B$ lifetime and mixing frequency and $\sin 2\beta$ measurement}

The precision measurement of the unitarity triangle angles, time-integrated and time-dependent CP violation (TDCPV) observables is a powerful test for the Standard Model (SM) consistency. 

The measurement of $B^0$ lifetime $\tau_{B^0}$ and mixing frequency  $\Delta m_d$ is performed in Ref.~\cite{Belle2:B0lifetime} using  ${B\to D^{(*)-}\pi^+}$ and ${B\to D^{(*)-}K^+}$ decays on the $190~\mathrm{fb}^{-1}$ Belle~II sample. 

The TDCPV analysis of $B^0\to J/\psi K_S$ decays is performed in Ref~\cite{Belle2:JpsiKS0} using the $190~\mathrm{fb}^{-1}$ Belle~II sample.  The proper-time difference of the two $B$ meson decays,  $\Delta t$, which distributes in
\begin{equation}
P(\Delta t)= e^{-|\Delta t|/\tau_{B^0}}/(4\tau_{B^0})\left\{1+q[A_{CP}\cos(\Delta m_d\Delta t)+S_{CP}\sin(\Delta m_d\Delta t)]\right\},
\end{equation}
is measured to extract the direct CP-violation ($A_{CP}$) and and mixed-induced CP-violation 
($S_{CP}$), where $q$ is the sign of the charge of the $b$ quark in the other side $B$ meson of the event. In the SM $A_{CP}\approx 0$ and $S_{CP}\approx \sin 2\beta$ are expected, where $\beta$ is the CKM matrix angle.

Both analyses results are not competitive,  but they validate the Belle~II performance in term of $\Delta t$ resolution and $B$-flavour tagging capabilities,  they serve as benchmark for TDCPV analyses and the low systematic uncertainties demonstrate the already good detector control.

\section{Time dependent CP violation analyses using gluonic penguins}

The $b\to q\overline qs$ transitions, which proceed via gluonic penguin decays, are suppressed in the SM and they have branching fractions in the $10^{-5}$-$10^{-6}$ ranges. The theory predictions are relatively precise and $A_{CP}$ and $S_{CP}$ measured from these decays can be used to test the SM and possibly access to the beyond the SM amplitudes. 
The reconstruction of these decays is challenging because of the presence of neutral particles in the final state, for the low yields and high backgrounds. However, these channels can be reconstructed only at Belle~II.

We report three TDCPV analysis performed on the $362~\mathrm{fb}^{-1}$ Belle~II sample: $B^0\to \phi(\to K^+K^-) K_S^0$, $B^0\to K_S^0K_S^0K_S^0$, $B^0\to K_S^0\pi^0$. In the first channel the main challenge is the discrimination of the non-resonant background from $B\to K^+K^-K_S^0$ decays, performed using an helicity angle fit. In the two latter channels, the challenge is the reconstruction of the $B^0$ vertex using only the displaced tracks from the $K_S^0$ decays.  
In all the channels a multi-dimensional fit is used to extract the signal events. The fits exploits $\Delta E=E_B^*-E_\mathrm{beam}^*$ (where $*$ means evaluated in the $\Upsilon(4S)$ frame, $E_B$ is the $B$ energy, $E_\mathrm{beam}$ is the beam energy),  $M_{\rm bc}=\sqrt{E^{*2}_\mathrm{beam}-p^{*2}_B} $,   $\Delta t$, and $O_{CS}$ (the output of the background suppression BDT) as fitting variables. Dedicated control channels are used to calibrate the fit variables.  The events with low $\Delta t$ quality information are used to constrain $A_{CP}$, fitted simultaneously to $S_{CP}$. The results for the three analyses are reported in Table~\ref{tab:gluonicPenguins}. The measured $S_{CP}$ values are compatible with the world averages, while $A_{CP}$ precision is on par with best measurements for the three channels~\cite{PDG}.

\begin{table}[!hbt]
\caption[]{Measured values of $A_{CP}$ and $S_{CP}$ for the three considered decay channels; the world average (w.a.) values are also reported as a reference~\cite{PDG}. The first uncertainty is statistical, the second systematic.}
\label{tab:gluonicPenguins}
\vspace{0.4cm}
\begin{center}
\begin{tabular}{|l c c c c |}
\hline
Decay                                                 & $A_{CP}^\mathrm{meas.}$           & $S_{CP}^\mathrm{meas.}$             & $A_{CP}^\mathrm{w.a.}$ & $S_{CP}^\mathrm{w.a.}$ \\
\hline
$B^0\to \phi K_S^0$                    &  $0.31\pm 0.20^{+0.05}_{-0.06}$ & $0.54\pm0.26^{+0.06}_{-0.08}$    & $-0.01\pm 0.14$                  & $0.59\pm0.14$ \\ 
$B^0\to K_S^0K_S^0K_S^0$     &  $0.07^{+0.15}_{-0.20}\pm 0.02$  &  $-1.37^{+0.35}_{-0.45}\pm 0.03$  & $0.15\pm0.12$                  & $-0.83\pm0.17$ \\
$B^0\to K_S^0\pi^0$                   &  $0.04^{+0.15}_{-0.14}\pm 0.05$  &  $0.75^{+0.20}_{-0.23}\pm 0.04$  & $0.0\pm0.13$                      & $0.58\pm0.17$ \\
 \hline
\end{tabular}
\end{center}
\end{table}

\section{Isospin sum rule}

The theoretical prediction of CP violation observables in the $B\to K\pi$ decays has typically large uncertainties. However in Ref.~\cite{Gronau:puzzle} an isospin sum rule is presented. It combines the $A_{CP}$ asymmetries from the $B\to K^+\pi^-$, $B^0\to K_S^0\pi^+$, $B^+\to K^+\pi^0$,  and $B^0\to K_S^0\pi^0$ to obtain the $I_{K\pi}$ observable, which is predicted to be 0 within 1\% in the SM. However, the experimental precision is at 10\% level~\cite{puzzle:exp}  driven by $A_{CP}^{K_S^0\pi^0}$.

\begin{figure}[!hbt]
\begin{minipage}{0.5\linewidth}
\centerline{\includegraphics[width=\linewidth,]{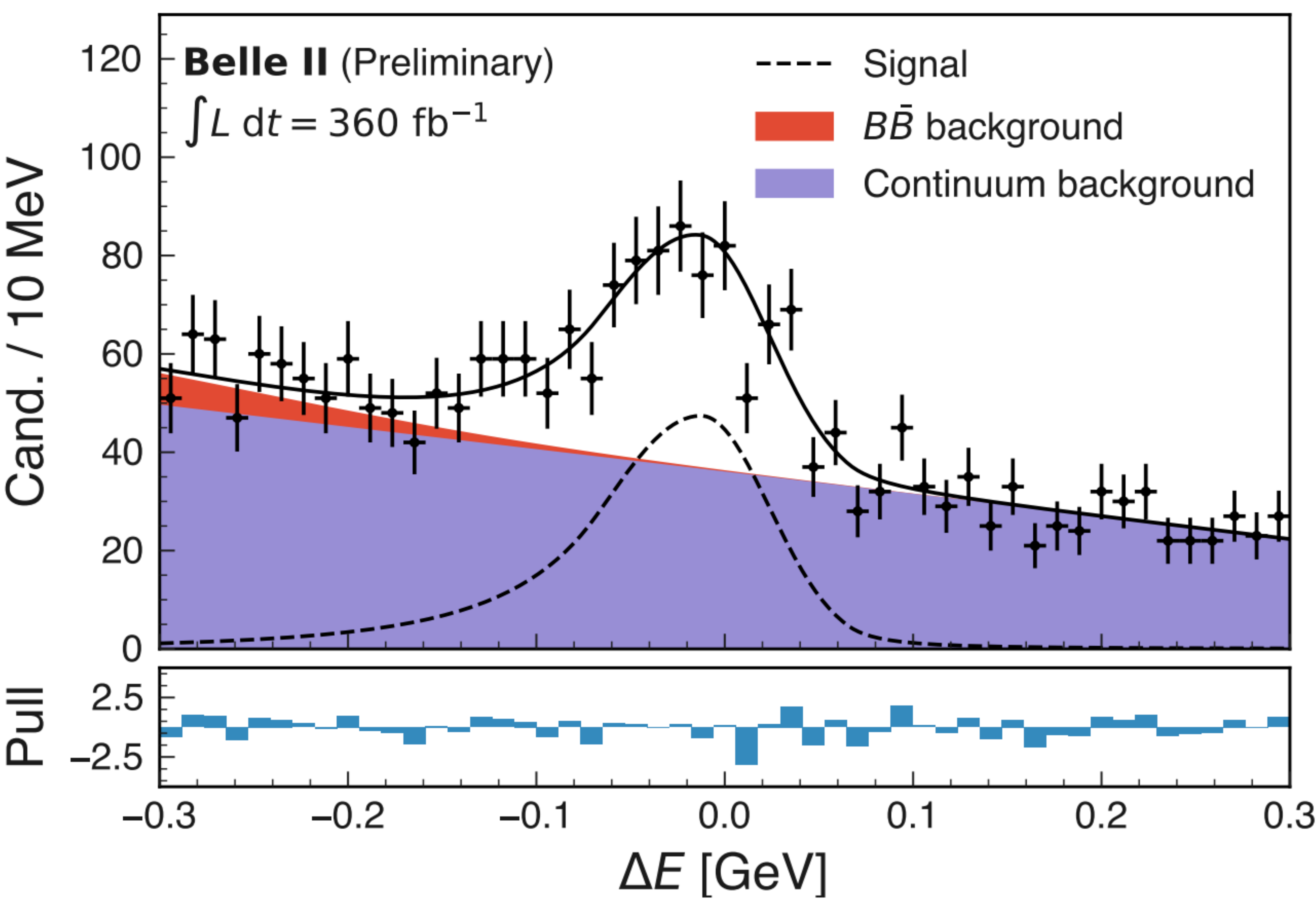}}
\end{minipage}
\hfill
\begin{minipage}{0.43\linewidth}
\centerline{\includegraphics[width=\linewidth,]{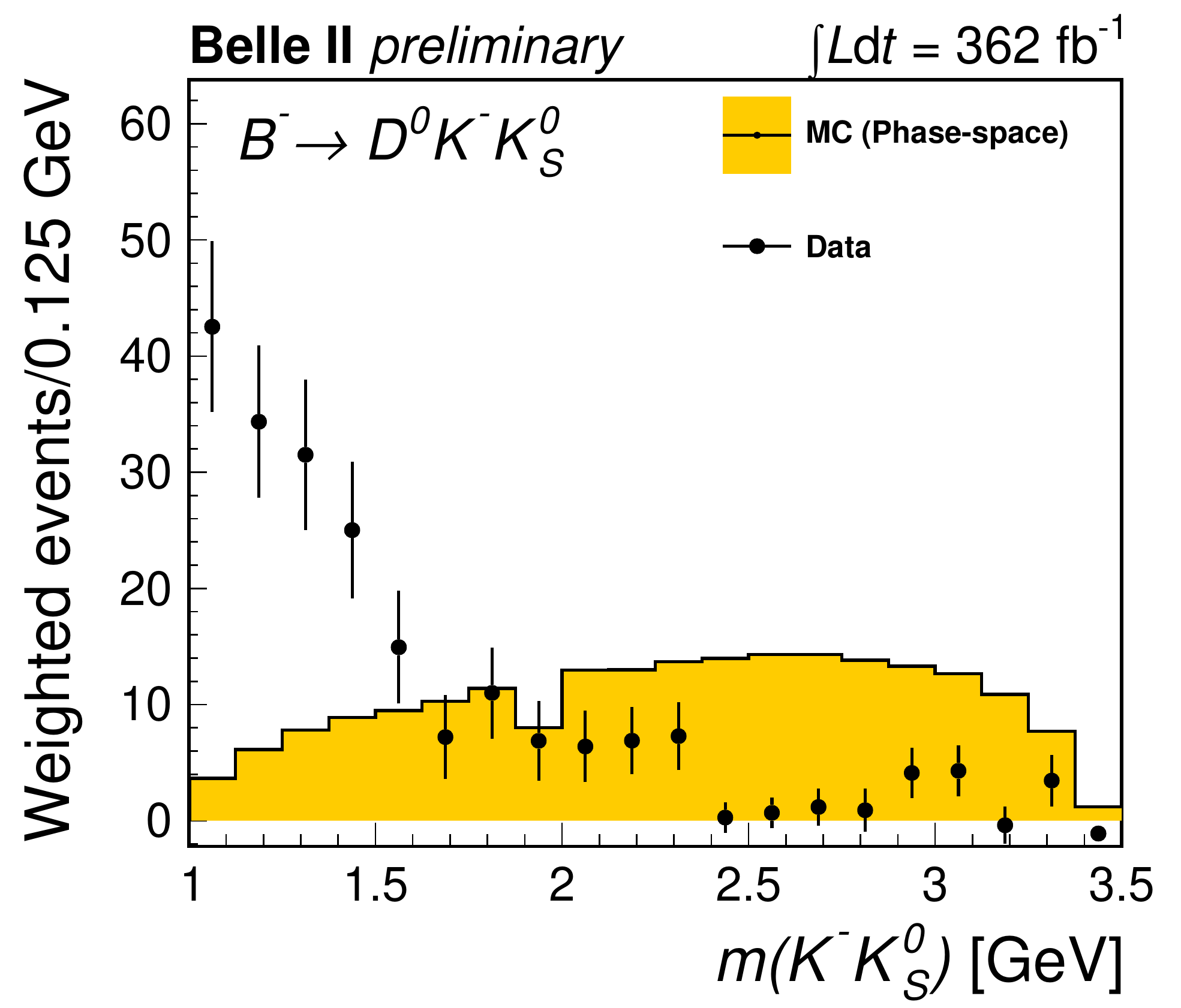}}
\end{minipage}
\hfill
\caption[]{Left:  $\Delta E$ distribution for $B^0\to K_S^0\pi^0$ decays with the fit projection overlaid. Right: $m(K^-K^{0}_S)$ distribution for $B^-\to D^0K^-K_S^0$ decays, comparing data and a phase-space simulation. }
\label{fig:Iso-DKK}
\end{figure}

We present the measurement of all the $B\to K\pi$ decays branching fraction and direct asymmetries $A_{CP}$ on the $362~\mathrm{fb}^{-1}$ Belle~II sample. The results are obtained from a time-integrated analysis and they are  competitive with the world best measurement~\cite{PDG}. In particular the analysis $B^0\to K_S^0\pi^0$  is combined with the time-dependent result presented in previous section to obtain the world best estimation of $A_{CP}^{K_S^0\pi^0}=-0.01\pm 0.12~~\mathrm{(stat.)}\pm 0.05~\mathrm{(syst.)}$, where the world average is $A_{CP}^{K_S^0\pi^0}=0.0\pm0.13$.  This result is exploited to asses the Belle~II estimation of $I_{K\pi}=-0.03\pm 0.13~\mathrm{(stat.)}\pm 0.05~\mathrm{(syst.)}$, competitive with the world average~\cite{PDG}. The fit projection is shown in Fig.~\ref{fig:Iso-DKK} (left).

\section{Measurement of $\gamma$ angle from $B\to DK$ decays}

The unitarity triangle $\gamma$ is a crucial observable to test the SM, since its theoretical prediction has very small uncertainties ($\Delta\gamma_\mathrm{theory}/\gamma\sim 10^{-7}$). The $\gamma$ angle is extracted from $B\to DK$ decays in the interference between the favoured $b\to c\overline us$ and suppressed $b\to u\overline cs$ transitions
\begin{equation}
\frac{A_\mathrm{sup}(B^-\to \overline D^0 K^-)}{A_\mathrm{fav}(B^-\to D^0 K^-)}=r_Be^{i(\delta_B-\gamma)},
\end{equation}
where $\delta_b$ is the strong phase difference and $r_B$ is the magnitude of the suppression. The measurement of $\gamma$ can be approached in multiple ways, according to the chosen $D$ final state. We present the measurement of $\gamma$ with GLS method~\cite{GLS} and GLW method~\cite{GLW1,GLW2}, while the Belle~II measurement using BPGGSZ method has been already published in Ref.~\cite{Belle2:gammaBPGGSZ}. 

The GLS method exploit the cabibbo-suppressed $D\to K_S^0K^\pm\pi^\mp$ decays to extract 7 CP observables (4 asymmetries, 3 branching ratios).  The measurement is performed using the combined $711~\mathrm{fb}^{-1}$ Belle and  $362~\mathrm{fb}^{-1}$ Belle~II samples.
The fit is performed both in the full $D$ phase space and in the interference-enriched $D\to KK^*$ region. The fit to the $\Delta E$ and $O_{CS}$ distributions is performed simultaneous in 8 categories (two $B$ charges, $B^\pm\to DK^\pm$ or $B^\pm\to D\pi^\pm$, and sample with the charge of the kaon from $D$ with the same or opposite charge of the $B^\pm$).  An external input to describe the $D$ decay parameters is needed to extract $\gamma$, for instance from Ref.~ \cite{CLEO:gamma}. The results set only a contraint to $\gamma$ and they are not competitive but should be considered within the framework of a future combination of the $\gamma$ extraction methods, exploiting combined Belle and Belle~II datasets. 

The GLW method exploits the CP eigenstates $D\to K^+K^-$ (CP-even) and $D\to K_S^0\pi^0$ (CP-odd), extracting $\gamma$ from  
$\mathcal R_{CP\pm}= 1+r_B^2\pm 2r_B\cos\delta_B\cos\gamma$, $\mathcal A_{CP\pm}=\pm 2 r_B\sin \delta_B\sin\gamma/\mathcal R_{CP\pm}$. 
The measurement is performed using the combined $711~\mathrm{fb}^{-1}$ Belle and  $189~\mathrm{fb}^{-1}$ Belle~II samples.
The $B^\pm\to D(\to K^+\pi^-)h$ decays, with $h=K^\pm, \pi^\pm$ is used to access to $D$ decay information. 
The fit to the $\Delta E$ and $O_{CS}$  distributions is performed simultaneous in 6 categories ($D\pi$ or $DK$, with $D$ decaying to CP-even, CP-odd or control channel modes). The result of the CP-even eigenstates and the $\gamma$ estimation are not competitive with the world average~\cite{PDG}. However, the CP-odd eigenstates are unique to Belle~II, and this measurement is the world best estimation of $A_{CP-}=(-16.7\pm 5.7~\mathrm{(stat.)}\pm 0.6~\mathrm{(syst.)})\%$, $\mathcal R_{CP-} =1.151\pm 0.074~\mathrm{(stat.)}\pm 0.019~\mathrm{(syst.)}$. In addition, $A_{CP+}$ and $A_{CP-}$ are has almost the same strength and they differ with a significance of $3.5\sigma$, as expected from the SM.

\section{Observation of $B\to D^{(*)} K^-K_S^0$ decays}

A large fraction of the Belle~II physics programme relying on $B$-tagging, a set of reconstruction techniques in which the partner $B$ meson, produced in association to the signal, is first reconstructed with specific channels to infer the properties of the signal using the $\Upsilon(4S)$ initial state constraint. 
The $B$-tagging is based on a set of multivariate classifier trained on the simulation~\cite{Keck:FEI}. 
However about 40\% of the $B$ exclusive branching fractions is not known and an improvement in the $B$ branching fractions knowledge can lead to an increase in $B$-tagging efficiency. 

The $B\to D^{(*)} K^{(*)} K^{(*)}$ sector is almost unexplored with only five measured modes~\cite{Belle:DKK}. We present the measurement of the branching fraction of the four $ B\to D^{(*)} K^- K^{0}_S$ decays  performed on the $362~\mathrm{fb}^{-1}$ Belle~II sample~\cite{Belle2:DKK}, including the first observation of $\overline B{}^0\to D^+K^-K_S^0$, $B^-\to D^{*0}K^-K_S^0$, $\overline B{}^0\to D^{*+}K^-K_S^0$ decays and an improvement in the precision of $\mathcal{B}(B^-\to D^0K^-K_S^0)$. The signals are extracted from a $\Delta E$ fits and they are efficiency-corrected as a function of $m(K^-K^{0}_S)$ invariant mass. We also show $m(K^-K^{0}_S)$ of the aforementioned decays, observing structures which are far from phase-space. An example is shown in Fig.~\ref{fig:Iso-DKK} (right).

\section{Fully inclusive $B\to X_s\gamma$ measurement}

The $B\to X_s\gamma$ transition is a radiative flavour changing neutral current, suppressed in the SM and particularly sensitive to new physics. We present the measurement the $B\to X_s\gamma$ branching ratio as a function of the photon energy in the range $1.8~\mathrm{GeV}<E_\gamma<2.7~\mathrm{GeV}$, where $X_s\gamma$ is the inclusive final state involving a photon and a strange hadron. The decays are reconstructed using the  hadronic $B$-tagging (i.e. reconstructing the partner $B$ with exclusive hadronic $B$ decays only), and requiring a $\gamma$ in the signal side. The background are suppressed using a BDT together with the information of the simulation for specific decays. The signal is extracted using a fit to $M_{\rm bc}$ distribution as a function of $E_\gamma$. The result is competitive with previous measurements performed with hadronic  $B$-tagging.

\section*{Acknowledgments}

This project has received funding from the European Union’s Horizon 2020 research and innovation programme under the ERC grant agreement No 819127.

\section*{References}

\bibliography{biblio.bib}

\end{document}